\documentclass[prd,twocolumn,showpacs]{revtex4}
\usepackage{graphicx}
\usepackage{dcolumn}
\usepackage{bm}
\usepackage{natbib}

\newcommand{\aap}{{Astron. Astrophys.}}
\newcommand{\apjl}{{Astrophys. J. Lett.}}

\newcommand{\aj}{{Astron. J.}}
\newcommand{\mnras}{{Mon. Not. R. Astron. Soc.}}

\newcommand{\nhat}{\hat{\bf n}}
\newcommand{\labc}{{\ell_\alpha\ell_\beta\ell_\gamma}}

\begin{document}

\title{Intrinsic alignment-lensing interference as a contaminant of cosmic
shear}

\author{Christopher M. Hirata}
\email{chirata@princeton.edu}

\author{Uro\v s Seljak}
\email{useljak@princeton.edu}
\affiliation{Dept. of Physics, Jadwin Hall, Princeton University, 
Princeton, New Jersey 08544, USA}

\date{accepted version August 18, 2004; corrected July 31, 2010}

\begin{abstract}
Cosmic shear surveys have great promise as tools for precision cosmology,
but can be subject to systematic errors including intrinsic
ellipticity correlations of the source galaxies.  The intrinsic alignments
are believed to be small for deep surveys, but this is based on intrinsic
and lensing distortions being uncorrelated.  Here we show that the
gravitational lensing shear and intrinsic shear need not be independent:
correlations between the tidal field and the intrinsic shear cause the
intrinsic shear of nearby galaxies to be correlated with the gravitational
shear acting on more distant galaxies.  We estimate the magnitude of this
effect for two simple intrinsic alignment models: one in which the galaxy
ellipticity is linearly related to the tidal field, and one in which it is
quadratic in the tidal field as suggested by tidal torque theory.  The
first model predicts a gravitational-intrinsic ($GI$) correlation that can
be much greater than the intrinsic-intrinsic ($II$) correlation for broad
redshift distributions, and that remains when galaxies pairs at similar
redshifts are rejected.  The second model, in its simplest form, predicts
no gravitational-intrinsic correlation.  In the first model and assuming a
normalization consistent with recently claimed detections of intrinsic
correlations we find that the $GI$ correlation term can exceed the usual
$II$ term by $>1$ order of magnitude and the intrinsic correlation
induced $B$-mode by 2 orders of magnitude. These interference effects
can suppress the lensing power spectrum for a single broad redshift
bin by of order $\sim 10\%$ at $z_s=1$ and $\sim 30\%$ at $z_s=0.5$.  We 
conclude that, depending on the intrinsic alignment model, the 
$GI$ correlation may be the dominant contaminant of the lensing signal and 
can even affect cross-spectra between widely separated bins.  We describe 
two ways to constrain this effect, one based on density-shear correlations 
and one based on scaling of the cross-correlation tomography signal with 
redshift.
\end{abstract}


\maketitle

\section{Introduction}
\label{sec:intro}

Weak gravitational lensing has attracted considerable interest recently as
a means of measuring the density perturbations in the universe at low
redshifts because it is directly sensitive to the matter distribution.  
It thus avoids the astrophysical complications involved in other means of
measuring the matter distribution such as galaxy clustering.  Recent first
detections of the cosmic shear autopower spectrum
\cite{2000A&A...358...30V, 2000MNRAS.318..625B, 2001ApJ...552L..85R,
2002ApJ...572...55H, 2002A&A...393..369V, 2003AJ....125.1014J,
2003MNRAS.341..100B, 2004astro.ph..4195M} have stimulated proposals for
more ambitious projects in the future, such as CFHTLS \footnote{URL: {\tt
http://www.cfht.hawaii.edu/Science/CFHLS/}}, Pan-STARRS \footnote{URL:
{\tt http://pan-starrs.ifa.hawaii.edu/public/}}, SNAP \footnote{URL: {\tt
http://snap.lbl.gov/}}, and LSST \footnote{URL: {\tt
http://www.lsst.org/}}, which aim for percent-level precision.  
Comparison of the shear power spectra at different redshifts and/or to the
primordial fluctuation amplitude derived from cosmic microwave background
(CMB) anisotropy measurements, can constrain the cosmological growth
factor and hence parameters such as the neutrino mass, amplitude of
fluctuations $\sigma_8$, and dark energy equation of state
\cite{1999ApJ...514L..65H, 1999ApJ...518....2E, 1997ApJ...488....1Z,
1997MNRAS.291L..33B, 2002PhRvD..66h3515H, 2002PhRvD..65b3003H}.

While the underlying physics of weak lensing is ``clean,'' the shear
measurements are subject to possible systematic errors including
incomplete correction for seeing and optical distortions, selection
effects, and noise-rectification biases \cite{2000ApJ...537..555K,
2001A&A...366..717E, 2001MNRAS.325.1065B, 2002AJ....123..583B,
2003MNRAS.343..459H, 2003astro.ph..5089V}, and their cosmological
interpretation relies on accurate knowledge of the redshift distribution
of the source galaxies. Another possible systematic error is intrinsic
(i.e. not lensing-induced) correlations among the ellipticities of
neighboring source galaxies \cite{2000ApJ...545..561C,
2000MNRAS.319..649H, 2000ApJ...532L...5L, 2000ApJ...543L.107P,
2001ApJ...555..106L, 2001MNRAS.320L...7C, 2001ApJ...559..552C,
2002ApJ...567L.111L, 2002MNRAS.335L..89J, 2002MNRAS.333..501B}, which
could arise if the galaxy ellipticities are affected by large-scale tidal
fields.  This systematic is particularly worrisome because it lies outside
the control of the observer, and is dependent upon the poorly understood
physics of galaxy formation.  One frequently proposed method to reduce
this contamination is to assign photometric redshifts to the source
galaxies, and then to down-weight or ignore pairs of galaxies at similar
redshifts when computing the shear correlation function or power spectrum
\cite{2002A&A...396..411K, 2003A&A...398...23K, 2003MNRAS.339..711H,
2004ApJ...601L...1T}; see Ref.~\cite{2004MNRAS.347..895H} for an
implementation.  The idea is that the pairs of galaxies widely separated
in redshift should have independent intrinsic ellipticities but should
have correlated gravitational shears induced by structures between the
observer and the more nearby source galaxy.

While cross-correlation of different redshift bins is expected to remove
spurious power due to the intrinsic alignment autocorrelation, there is
another, more subtle effect by which even the cross-correlations can be
contaminated from intrinsic alignments.  If the ellipticities of galaxies
are correlated with the tidal quadrupole field in which the galaxies form,
then the intrinsic ellipticity of a nearby source galaxy will be
correlated with the lensing shear acting on a more distant source galaxy.  
This leads to a nonzero cross-correlation between the intrinsic
ellipticity and the gravitational lensing shear.  Cross-correlations
between the shear measurements at widely different redshifts are actually
more contaminated by this effect than shear autocorrelations computed
using only galaxies in a narrow redshift slice, because the radial
separation is necessary in order for the tidal field around the nearby
galaxy to lens the more distant galaxy.  This intrinsic-lensing
correlation will also affect attempts to cross-correlate cosmic shear
surveys of galaxies with lensing of the CMB \cite{2002ApJ...574..566H,
2002PhRvD..65b3003H, 2003PhRvD..67d3001H} because the CMB is also lensed
by the tidal field surrounding the source galaxies.

This paper is organized as follows.  In Sec.~\ref{sec:power}, we formally
express the $E$- and $B$-mode shear power spectra in terms of the
background cosmology, power spectra of matter density and intrinsic shear,
and matter-shear cross-spectrum.  In Sec.~\ref{sec:predict} we consider
two crude models of intrinsic alignments and calculate their predicted
contribution to the shear power spectrum.  We discuss methods to assess
and/or remove the contamination in Sec.~\ref{sec:methods}, and we conclude
in Sec.~\ref{sec:discussion}.

\section{Shear power spectra}
\label{sec:power}

\begin{figure}
\includegraphics[width=3in]{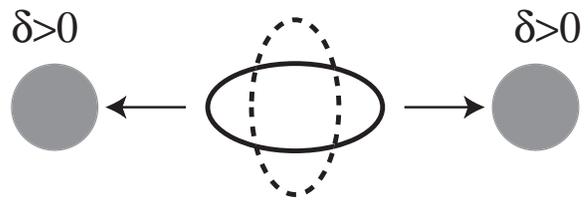}
\caption{\label{fig:shear}The effect of the density-intrinsic shear
correlation on the shear power spectrum.  Density fluctuations
in the nearby plane (gray masses) induce a tidal field (arrows).  A source
galaxy in a more distant plane (dashed ellipse) is
gravitationally sheared tangentially to these masses.  If the intrinsic
shears of galaxies in the nearby plane (solid
ellipse) are aligned with the stretching axis of the tidal field, then
this results in an anti-correlation between the shears of
galaxies at different redshifts, i.e. $C_\ell^{EE,GI}<0$.  (The opposite
case, $C_\ell^{EE,GI}>0$, results if galaxies are
preferentially aligned with the compressing axis of the tidal field.)}
\end{figure}

Before developing the formalism let us describe a simple example of the
effect, shown in Fig. \ref{fig:shear}. The tidal field may lead to a
stretching of the galaxy shape in the direction of the tidal field.  
Gravitational shearing of a background source leads to stretching of the
galaxy a in perpendicular direction. As a result, the lensing effect will
be partially cancelled by the intrinsic alignemnt effect and the two
effects are coherent, as they depend on the same underlying density field.

Weak gravitational lensing by large-scale structure is detectable through
its shearing of distant ``source'' galaxies.  To lowest order, the shear
of a galaxy $i$ can be broken down into a gravitational and an intrinsic
shear contribution: $\gamma_i = \gamma^G_i + \gamma^I_i$.  The
gravitational shear is well known and is equal to:
\begin{equation}
(\gamma^G_{i+}, \gamma^G_{i\times}) = \partial^{-2}
\int_0^\infty W(\chi,\chi_i) (\partial^2_x-\partial^2_y, 
2\partial_x\partial_y) \delta(\chi\nhat_i) d\chi,
\label{eq:gamma-g}
\end{equation}
where the $\partial$ derivatives are takes with respect to angular
position (i.e. have units of radians$^{-1}$), $\partial^{-2}$ is the
associated inverse Laplacian, $\nhat_i$ is the angular position of galaxy
$i$, $\delta(\chi\nhat)$ is the fractional density perturbation at
distance $\chi$ in direction $\nhat$, and the lensing window function is:
\begin{equation}
W(\chi,\chi_i) = 
{3\over 2}\Omega_mH_0^2(1+z)\sin^2_K\chi\,(\cot_K\chi-\cot_K\chi_i)
\label{eq:grav-window}
\end{equation}
for $\chi<\chi_i$ and $0$ otherwise.  Here $\sin_K$ and $\cot_K$ are the 
modified trigonometric functions, i.e.
\begin{equation}
\sin_K\chi = \left\{ \begin{array}{lcl}
K^{-1/2}\sin (K^{1/2}\chi) & & K>0 \\
\chi & & K=0 \\
|K|^{-1/2}\sinh (|K|^{1/2}\chi) & & K<0 \end{array} \right.,
\end{equation}
$\cot_K\chi = {d\over d\chi}\ln \sin_K\chi$, and $K$ is the 
spatial curvature of the universe.

Now we imagine that a set of source galaxies in redshift slice $\alpha$
with comoving distance distribution $f_\alpha(\chi)$ are observed.  The
gravitational contribution to a shear Fourier mode is:
\begin{equation}
\gamma^G_{\bf l}(\alpha) = (\cos 2\phi_{\bf l},\sin 2\phi_{\bf l}) 
\int_0^\infty W_\alpha(\chi) \delta_{\bf l}(\chi) d\chi,
\label{eq:gammal}
\end{equation}
where the integrated window function is:
\begin{equation}
W_\alpha(\chi) = \int_0^\infty f_\alpha(\chi') W(\chi,\chi') d\chi'. 
\label{eq:wint}
\end{equation}

The intrinsic alignment contribution at a point is given by
\begin{equation}
\gamma^I(\nhat,\alpha) = \int_0^\infty 
f_\alpha(\chi)\tilde\gamma^I(\chi\nhat)d\chi,
\label{eq:gi}
\end{equation}
where the density-weighted intrinsic shear
$\tilde\gamma^I=(1+\delta_g)\gamma^I$ is computed from the fractional
overdensity of galaxies $\delta_g$ and average intrinsic shear of galaxies
$\gamma^I$.  A density weighting is technically necessary in
Eq.~(\ref{eq:wint}), which makes $W_\alpha(\chi)$ slightly dependent on
angular position.  On sub-arcminute scales where the fluctuations in
$\delta_g$ are large this results in production of $B$-modes in the
lensing shear \cite{2002A&A...389..729S}; on larger scales the effect is
unimportant. In contrast, intrinsically aligned pairs of galaxies tend to
be close to each other where $\delta_g\ge 1$, and hence the factor of
$1+\delta_g$ in Eq.~(\ref{eq:gi}) cannot be safely neglected except on
very large scales.

The $E$-mode shear cross-spectrum between two redshift slices can be
broken down into gravitational lensing ($GG$), intrinsic alignment ($II$),
and interference ($GI$)  terms:
\begin{equation}
C^{EE}_\ell(\alpha\beta) = C^{EE,GG}_\ell(\alpha\beta) + 
C^{EE,II}_\ell(\alpha\beta) + C^{EE,GI}_\ell(\alpha\beta).
\label{eq:cesum}
\end{equation}
(The $B$-mode shear cross-spectrum is similar, but contains only an $II$
term since there is no gravitational contribution to the $B$-mode shear.)  
The gravitational lensing contribution to the shear power spectrum can be
obtained on small scales ($\ell\gg 1$) by Limber integration
\cite{1992ApJ...388..272K}, which computes the power spectrum of a
radially projected quantity.  In this case we integrate
Eq.~(\ref{eq:gammal}) to get
\begin{equation}
C^{EE,GG}_\ell(\alpha\beta) = \int_0^\infty {W_\alpha(\chi) 
W_\beta(\chi)\over\sin_K^2\chi} P_\delta(k={\ell\over\sin_K\chi};\chi) 
d\chi.
\end{equation}
The intrinsic alignment contribution is most easily computed by Limber
integration of Eq.~(\ref{eq:gi}):
\begin{equation}
C^{EE,II}_\ell(\alpha\beta) = \int_0^\infty {f_\alpha(\chi) 
f_\beta(\chi)\over \sin_K^2\chi} 
P_{\tilde\gamma^I}^{EE}(k={\ell\over\sin_K\chi};\chi) d\chi,
\label{eq:ceii}
\end{equation}
where the projected power spectrum of the intrinsic alignments is:
\begin{eqnarray}
P_{\tilde\gamma^I}^{EE}(k;\chi) \!\! &=& \pi \int\int
\bigl\{
\langle \tilde\gamma^I_+(0) \tilde\gamma^I_+(x) \rangle 
[J_0(kx_\perp)+J_4(kx_\perp)]
\nonumber \\ &&
+ \langle \tilde\gamma^I_\times(0)
\tilde\gamma^I_\times(x) \rangle [J_0(kx_\perp)-J_4(kx_\perp)]
\bigr\}
\nonumber \\ && \times x_\perp dx_\perp dx_{||},
\label{eq:pe}
\end{eqnarray}
where the $\langle\rangle$ are two-point correlation functions of
$\tilde\gamma$ with two points separated by comoving distance $x_{||}$ in
the radial direction and $x_\perp$ in the perpendicular direction.  The
$+$ and $\times$ components of the shear are measured along the direction
of separation ${\bf x}_\perp$ and at 45 degrees to this direction,
respectively.  Note that in general the two-point correlation functions
can depend on $x_\perp$ and $x_{||}$ independently (i.e. they are not just
functions of the separation $\sqrt{x_\perp^2+x_{||}^2}$) since the shear
depends on viewing geometry.  To get the $B$-mode power, switch the $+$
and $\times$ labels in Eq.~(\ref{eq:pe}).

Given models for intrinsic alignment and galaxy clustering,
Eqs.~(\ref{eq:ceii}) and (\ref{eq:pe})  are sufficient to compute the pure
intrinsic alignment contribution to the shear cross-spectrum.  However,
note that as a second-order statistic, Eq.~(\ref{eq:cesum}) also contains
interference terms due to correlation of the gravitational shear with
intrinsic alignments.  These have normally been ignored because the
gravitational shear acting on a particular galaxy is determined by the
integrated tidal quadrupole along the line of sight, whereas the intrinsic
shear is expected to be determined by tidal fields in the vicinity of the
galaxy.  But in cosmic shear studies, we correlate the shears of two
galaxies whose redshifts may be different.  In this case, it is plausible
that the more nearby galaxy is intrinsically aligned by a quadrupolar
tidal field that also lenses the more distant galaxy.  Therefore we must
consider the interference terms.  These are given by
\begin{eqnarray}
C^{EE,GI}_\ell(\alpha\beta) &=& \int_0^\infty {W_\alpha(\chi) 
f_\beta(\chi) + W_\beta(\chi)f_\alpha(\chi) \over\sin_K^2\chi} 
\nonumber \\ && \times
P_{\delta,\tilde\gamma^I}({\ell\over\sin_K\chi}) d\chi,
\label{eq:cegi}
\end{eqnarray}
where the 3-dimensional cross-spectrum $P_{\delta,\tilde\gamma^I}$ is 
related to the density-shear correlation function via
\begin{equation}
P_{\delta,\tilde\gamma^I}(k;\chi) = -2\pi \int\int \langle \delta(0) 
\tilde\gamma^I_+(x) \rangle J_2(kx_\perp) x_\perp dx_\perp dx_{||}.
\label{eq:pegi}
\end{equation}

\section{Models}
\label{sec:predict}

We now compute the predicted level of contamination of weak lensing
surveys from the ``interference'' terms, Eq.~(\ref{eq:cegi}).  We
consider three cases: first the case of galaxies whose mean
ellipticities are linear functions of the underlying tidal field;
and secondly, the case of galaxies whose mean ellipticities are
quadratic functions of the tidal field.  In all of our numerical
results below, we have taken the bias to be $b_g=1$, used the
best-fit scale-invariant ($n_s=1$) flat $\Lambda$CDM model of
Ref.~\cite{2004PhRvD..69j3501T} ($\sigma_8=0.966$,
$\Omega_b=0.0475$, $\Omega_m=0.293$, $H_0=70.8$~km/s/Mpc), and used
the adiabatic CDM+baryon transfer function of
Ref.~\cite{1998ApJ...496..605E}.  The gravitational-gravitational
shear power spectra include the nonlinear correction of
Ref.~\cite{1996MNRAS.280L..19P}.

The first (linear) model has usually been used for elliptical
galaxies (since gravitational collapse in a general tidal field is
expected to result in a triaxial halo aligned with the principal
axes of the tidal field) and the second (quadratic) model for
spirals (since the orientation of these is believed to be
determined by angular momentum acquired during gravitational
collapse, which requires one tidal quadrupole to supply the torque
and another tidal quadrupole to be torqued). The models presented
here are ``toy'' models in the sense that their theoretical
motivation does not apply to highly nonlinear scales, and even on
larger scales there is little observational basis for models of
intrinsic galaxy ellipticity alignments. An alternative to analytic
models is to estimate intrinsic alignments by calculating the
ellipticities or angular momenta of dark matter haloes in $N$-body
simulations \cite{2000ApJ...532L...5L, 2000MNRAS.319..649H,
2000ApJ...545..561C, 2002MNRAS.335L..89J}, however one should keep
in mind that there can be misalignment between the dark matter halo
and the galaxy it contains \cite{2002ApJ...576...21V}, and so
intrinsic alignment results based on $N$-body simulations are not
definitive.  In principle it would be possible to derive
predictions for intrinsic correlations from a halo model, however
this would require adding a prescription for the galaxy alignment
to the current halo models.  Such a prescription would likely have
to come from an analytic or $N$-body model.  Undoubtedly there is
much room for improvement in both the theory -- which must
ultimately address the complications of galaxy formation, mergers,
etc. -- and the observations, which will be necessary to measure or
constrain the intrinsic alignment signal at levels suitable for
precision cosmology.

\subsection{Linear alignment model}
\label{ss:m1}

A simple model for the ellipticities of elliptical galaxies was proposed
by \cite{2001MNRAS.320L...7C}.  The intrinsic shear of the galaxy is
assumed to follow the linear relation
\begin{equation}
\gamma^I = -{C_1\over 4\pi G}(\nabla_x^2-\nabla_y^2, 
2\nabla_x\nabla_y){\cal S}[\Psi_P],
\label{eq:model1}
\end{equation}
where $\Psi_P$ is the Newtonian potential at the time of galaxy formation
(assumed to be during matter domination), ${\cal S}$ is a smoothing filter
that cuts off fluctuations on galactic scales, and $\nabla$ is a comoving
derivative (as opposed to $\partial$ which is a 2-dimensional derivative
on the unit sphere with units of radians$^{-1}$).  We have taken ${\cal
S}$ to be a simple cutoff in Fourier space at $k_{\rm cutoff}=1h$/Mpc.  
Here $C_1$ is a normalization constant (note that $C_1>0$ if the galaxy is
aligned along the ``stretching'' axis of the tidal field).  The original
motivation for Eq.~(\ref{eq:model1}) was the assumption that galaxies are
homologous with their haloes, and that the halo ellipticity is perturbed
by the local tidal field produced by large scale structure
\cite{2001MNRAS.320L...7C}.  On sufficiently large scales, this relation
can also be motivated by arguments analogous to linear biasing theory for
galaxies \cite{2002astro.ph..5512H}: the large-scale correlations in the
intrinsic shear field must be determined by the large-scale potential
fluctuations; if these large-scale potential fluctuations are sufficiently
small, then the intrinsic shear field should be a linear and local
function of the potential $\Psi_P$ (which is $\propto\Psi$ in the linear
regime); the only linear, local functions of $\Psi_P$ with quadrupole
symmetry are Eq.~(\ref{eq:model1}) and derivatives thereof; and on large
scales we expect higher-derivative terms such as
$\nabla^2(\nabla_x^2-\nabla_y^2,2\nabla_x\nabla_y){\cal S}[\Psi_P]$ to be
negligible.

The primordial potential is related to the linear density field via:
\begin{equation}
\Psi_P({\bf k}) = -4\pi G{\bar\rho(z)\over \bar 
D(z)}a^2k^{-2}\delta_{lin}({\bf k}),
\end{equation}
where $\bar\rho(z)$ is the mean density of the universe, $\bar D(z)\propto
(1+z)D(z)$ is the re-scaled growth factor normalized to unity during
matter domination, and $G$ is the Newtonian gravitational constant. On
linear scales, we have $\delta_g = b_g\delta_{lin}$,
$\delta=\delta_{lin}$, and hence the weighted intrinsic shear is:
\begin{eqnarray}
\tilde\gamma({\bf k}) &=& {C_1\bar\rho\over\bar D}a^2 \int 
{(k_{2x}^2-k_{2y}^2, 2k_{2x}k_{2y})\over k_2^2} \delta_{lin}({\bf k}_2) \Bigl[ 
\delta^{(3)}({\bf k}_1) 
\nonumber \\ &&
+ {b_g\over (2\pi)^3} \delta_{lin}({\bf k}_1)\Bigr] d^3{\bf k}_1,
\label{eq:gtilde}
\end{eqnarray}
where ${\bf k}_2\equiv{\bf k}-{\bf k}_1$ and we have chosen the wave
vector ${\bf k}$ to lie on the $x$-axis.  (We are only interested in modes
with ${\bf k}$ perpendicular to the line of sight.)  The power spectrum of
$\tilde\gamma$ is:
\begin{eqnarray}
P^{EE}_{\tilde\gamma^I}(k) &=& {C_1^2\bar\rho^2\over\bar D^2}a^4 \Bigl\{ 
P_\delta^{lin}(k)
+ b_g^2\int [f_E({\bf k}_2) + f_E({\bf k}_1) ]
\nonumber \\ && \times
f_E({\bf k}_2) { P_\delta^{lin}(k_1)  P_\delta^{lin}(k_2) \over (2\pi)^3}
d^3{\bf k}_1 \Bigr\},
\label{eq:emode}
\end{eqnarray}
where $f_E({\bf w}) = (w_{x}^2-w_{y}^2)/w^2$.  To get the $B$-mode power 
spectrum, we replace $f_E$ with $f_B({\bf w}) = 2w_xw_y/w^2$:
\begin{eqnarray}
P^{BB}_{\tilde\gamma^I}(k) &=& {C_1^2\bar\rho^2\over \bar D^2}a^4 b_g^2\int 
[f_B({\bf k}_2) + f_B({\bf k}_1) ]
\nonumber \\ && \times
f_B({\bf k}_2) { P_\delta^{lin}(k_1)  
P_\delta^{lin}(k_2) \over (2\pi)^3} d^3{\bf k}_1.
\label{eq:bmode}
\end{eqnarray}
Note that the $B$-mode intrinsic alignment power spectrum contains no
$O(P^{lin}_\delta)$ contribution.  This is because the tidal quadrupole
field is a pure $E$-mode (this is even true in the presence of nonlinear
evolution), and produces only an $E$-mode pattern of intrinsic alignments
if a linear model of galaxy ellipticities such as Eq.~(\ref{eq:model1})
applies.  Galaxy clustering modulates this field according to the galaxy
distribution, and thereby transfers some power into $B$-modes.

The cross-power of the matter density and weighted shear is then found
from Eq.~(\ref{eq:gtilde}):
\begin{equation}
P_{\delta,\tilde\gamma^I}(k) = -{C_1\bar\rho\over\bar D}a^2 P_\delta^{lin}(k).
\label{eq:cross}
\end{equation}

\subsection{Quadratic alignment model}
\label{ss:m2}

The apparent ellipticity of a spiral galaxy is determined principally by
the orientation (and hence angular momentum) of its disk. This angular
momentum is believed to comes from external tidal fields perturbing the
collapsing galaxy to form an anisotropic moment of inertia, which allows
the galaxy to be ``spun up'' by a tidal quadrupole.  In this case, the
angular momentum vector ${\bf L}$ of the galaxy acquires an anisotropic
probability distribution, and leads to a mean intrinsic ellipticity.  
This mean ellipticity vanishes to first order in the tidal field since a
tidal field is required both to produce the anisotropic moment of inertia
and then to apply a torque.  The second-order contribution is
\cite{2001MNRAS.320L...7C}
\begin{equation}
\gamma^I = C_2 (T_{x\mu}^2 - T_{y\mu}^2, 2T_{x\mu}T_{y\mu}),
\label{eq:model2}
\end{equation}
where the tidal tensor is
\begin{equation}
T_{\mu\nu} = {1\over 4\pi G}\left( \nabla_\mu\nabla_\nu - {1\over 
3}\delta_{\mu\nu}\nabla^2 \right) {\cal S}[\Psi_P].
\end{equation}
(Since the tidal shears $T$ are spin 2, we recall from the theory of
addition of angular momenta that Eq.~\ref{eq:model2} is the most general
quadratic function of $T$ with spin 2.)  The resulting density-weighted
intrinsic alignment is
\begin{eqnarray}
\tilde\gamma^I_E({\bf k}) &=& {C_2\bar\rho^2\over (2\pi)^3\bar D^2}a^4\int 
h_E(\hat{\bf k}'_1,\hat{\bf k}'_2)
\delta_{lin}({\bf k}'_1)\delta_{lin}({\bf k}'_2)
\nonumber \\ && \times
\left[ \delta^{(3)}({\bf k}'_3) + {b_g\over (2\pi)^3}\delta_{lin}({\bf 
k}'_3)\right]\, d^3{\bf k}'_1\, d^3{\bf k}'_2,
\end{eqnarray}
where we have retained the convention that ${\bf k}$ is chosen to lie on
the $x$ axis, set $\hat{\bf k}_a={\bf k}_a/|k_a|$ and ${\bf k}'_3={\bf
k}-{\bf k}'_1-{\bf k}'_2$, and defined the $h$-function by
$h_E=h_{xx}-h_{yy}$ with
\begin{equation}
h_{\lambda\mu}(\hat{\bf u},\hat{\bf v}) = \left( \hat u_\mu \hat u_\nu - 
{1\over 3}\delta_{\mu\nu} \right)
\left( \hat v_\lambda \hat v_\nu - {1\over 3}\delta_{\lambda\nu} \right).
\end{equation}
The $B$-mode intrinsic alignment is obtained by replacing $h_E$ with
$h_B=2h_{xy}$.  (See Ref.~\cite{2002MNRAS.332..788M} for a similar 
calculation without the density-weighting $1+\delta_g$.)

The predicted intrinsic alignment power spectra $P_{\tilde\gamma^I}(k)$
from this model can be derived by the same methods as used in
Sec.~\ref{ss:m1}, although since $\gamma^I$ is second-order in $\Psi_P$
(and hence $\delta_{lin}$) there are more Feynman diagrams that
contribute.  The result is
\begin{eqnarray}
P^{EE}_{\tilde\gamma^I}(k) \!\! &=& \!\! {C_2^2\bar\rho^4\over\bar D^4}a^8\Bigl\{
2 \int [h_E(\hat{\bf k}_1,\hat{\bf k}_2)]^2 {P_\delta^{lin}(k_1) 
P_\delta^{lin}(k_2)\over (2\pi)^3} d^3{\bf k}_1
\nonumber \\ && \!\!
+ {2\over 3}b_g^2 \int \bigl[ h_E(\hat{\bf k}'_1,\hat{\bf k}'_2) + 
h_E(\hat{\bf k}'_2,\hat{\bf k}'_3)
\nonumber \\ && \!\!
+ h_E(\hat{\bf k}'_3,\hat{\bf k}'_1)\bigr]^2 
\nonumber \\ && \!\! \times
{P_\delta^{lin}(k'_1) P_\delta^{lin}(k'_2) P_\delta^{lin}(k'_3)\over (2\pi)^6}
d^3{\bf k}'_1 d^3{\bf k}'_2 \Bigr\}.
\label{eq:pe.quad}
\end{eqnarray}
In Eq.~(\ref{eq:pe.quad}) we have defined ${\bf k}_2={\bf k}-{\bf k}_1$
and ${\bf k}'_3={\bf k}-{\bf k}'_1-{\bf k}'_2$.

In the simple version of the quadratic-alignment model presented here,
there is no density-shear correlation, i.e. $\delta_{lin}$ and
$(1+\delta_g)\gamma^I$ have zero cross-correlation assuming Gaussian
$\delta_{lin}$, linear biasing $\delta_g=b_g\delta_{lin}$, and linear
evolution of the density field.  We can see this as follows: since the
three-point functions of $\delta_{lin}$ vanish, this correlation is equal
to that of $\delta_{lin}$ and $\delta_g\gamma^I$.  Of the contributing
diagrams, those connecting $\delta_{lin}$ to the $\delta_g$ vanish because
$\langle\gamma^I\rangle=0$, hence we only consider the diagrams connecting
$\delta_{lin}$ to one of the two tidal quadrupoles $T_{\mu\nu}$ that make
up $\gamma^I$.  But these vanish by symmetry since they must connect
$\delta_g$, a scalar, to another tidal quadrupole $T_{\mu\nu}$ evaluated
at the same point.  This fact was first noted in the context of intrinsic
alignment contamination of galaxy-galaxy lensing by
Ref.~\cite{2002astro.ph..5512H}, which also pointed out that there can be
nonzero cross-power $P_{\delta,\tilde\gamma^I}(k)$ if we consider
nonlinear density evolution.  We will not consider these more complicated
models here.

\subsection{Normalization}
\label{ss:norm}

We have normalized the intrinsic alignment amplitude $C_1$ for the linear
alignment model to match the shear autopower observed by
Ref.~\cite{2002MNRAS.333..501B} in the SuperCOSMOS data
\cite{2001MNRAS.326.1279H}.  Ref.~\cite{2002MNRAS.333..501B} gives the
ellipticity variance $\sigma^2_e(\theta)$ in square cells of side length
$\theta$ instead of the shear power spectra; these are related by
\begin{equation}
\sigma^2_e(\theta) = {\cal R}^2\int j_0^2\left({\ell_x\theta\over 
2}\right) j_0^2\left({\ell_y\theta\over 2}\right)
( C_\ell^{EE} + C_\ell^{BB} ){d^2{\bf l}\over 4\pi^2},
\end{equation}
where $j_0(u)=\sin u/u$ and ${\cal R}=de/d\gamma\approx 2$ is the
shear-to-ellipticity conversion factor.  We have assumed a parameterized
redshift distribution of the form
\begin{equation}
{dn\over dz}\propto z^2 e^{-(1.4z/z_m)^{1.5}},
\label{eq:mg}
\end{equation}
where the median redshift for SuperCOSMOS is $z_m=0.1$
\cite{2002MNRAS.333..501B}.  The SuperCOSMOS ellipticity variances are an
order of magnitude greater than the weak lensing prediction and hence must
be dominated by intrinsic alignments and/or observational systematics; we
have assumed here that the intrinsic alignments dominate, which provides
an upper limit if observational systematics are not negligible.  We have
normalized to the largest-scale observation given by
Ref.~\cite{2002MNRAS.333..501B}, where Eq.~(\ref{eq:model1}) is best
motivated, and used the ellipticity covariance between the $b_J$ and $R$
bandpasses as this is less sensitive to systematics than the variances;
this yields $\sigma^2_e=1.6\times 10^{-5}$ at $\theta = 93.25$~arcmin.

For the linear intrinsic alignment model there is a discrete degeneracy in
$C_1$ since the sign cannot be determined from the ellipticity variance
alone; we have used in the plots the positive normalization $C_1>0$
because of its physical motivation (stretching of the galaxy along the
tidal field).

\subsection{Results}
\label{sec:results}

\begin{figure*}
\includegraphics[angle=-90, width=6.5in]{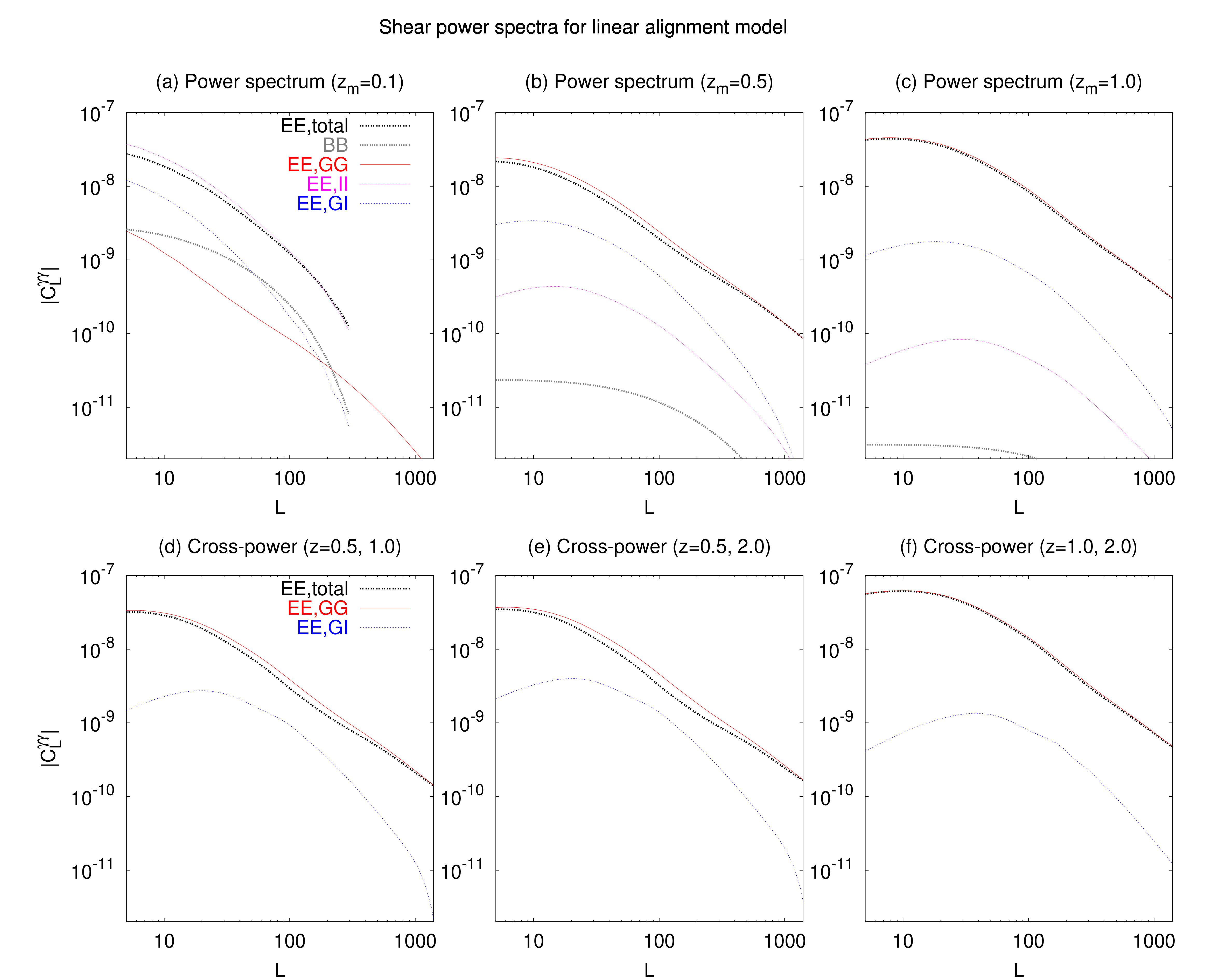}
\caption{\label{fig:broad}The gravitational ($GG$), intrinsic-alignment
($II$), and gravitational-intrinsic correlation ($GI$)  contributions to
the shear power spectrum, for the linear alignment model.  The $B$-mode
shear power spectrum is labeled ``B'' and has only an intrinsic-intrinsic
contribution.  The cross-term ($GI$) is negative for this model, so we
have plotted its absolute values on the log scale.  We show results for
(a) the shear power spectrum for a survey with redshift distribution of
Eq.~(\ref{eq:mg}) with $z_m=0.1$; (b) and (c) similar for for $z_m=0.5$
and $1.0$, respectively; (d) the shear cross-power between redshifts
$z=0.5$ and $z=1.0$ (the slices have Gaussian redshift distributions with
width $\sigma_z=0.1$); (e) the shear cross-power between redshifts $z=0.5$
and $z=2.0$; and (f) the shear cross-power between redshifts $z=1.0$ and
$z=2.0$.  The intrinsic alignment amplitude $C_1$ is normalized to
SuperCOSMOS.  Panel (a) is cut off at $\ell=300$, roughly the smoothing
scale used for the intrinsic alignment calculation, since the model does
not make sense on smaller scales.}
\end{figure*}

In Fig.~\ref{fig:broad}, we have shown the estimated intrinsic alignment
contamination of the shear power spectrum for the linear alignment model.  
Panels (a--c) shows the contributions to the $E$-mode shear power spectrum
assuming a source redshift distribution given by Eq.~(\ref{eq:mg}) with
median redshifts $z_m=0.1$, $0.5$, and $1.0$.  In panels (d), (e), and (f)
we have shown the intrinsic alignment contributions to the cross-spectra
between different redshift bins; only the gravitational-intrinsic
correlation is nonzero since the redshift distributions do not overlap.  
For these panels we take the redshift distributions to be Gaussian with
$\sigma_z=0.1$ (which is much less than the separation of the bins).  The
contamination $C^{GI}_\ell$ is only weakly dependent on $\sigma_z$ because
the integral in Eq.~(\ref{eq:cegi}) receives a contribution from a range
of radii $\Delta\chi\propto\sigma_z$, whereas the integrand contains a
factor of $f_\beta(\chi)$, whose peak value scales $\propto\sigma_z^{-1}$.

The most striking result from Fig.~\ref{fig:broad} is that the
gravitational-intrinsic contribution dominates the contamination of the
shear power spectrum for the broad redshift distribution
(Fig.~\ref{fig:broad}c). Even if the intrinsic-intrinsic shear
correlations are small (in this case they are $0.5\%$ of the lensing
signal for the broad redshift distribution with $z_m=1.0$ at $\ell=500$),
the gravitational-intrinsic contamination to the cross-spectra can still
be large ($\sim 5\%$ at $\ell=500$).  This contamination is especially
pronounced for the cross-spectrum between the widely separated bins at
$z=0.5$ and $2.0$ (Fig.~\ref{fig:broad}e): here
$|C_\ell^{GI}/C_\ell^{GG}|$ can be as large as $\sim 30\%$.

As noted above, there is no correlation between the gravitational and
intrinsic shears in the quadratic alignment model.  Thus there is only
intrinsic alignment contamination if the redshift distributions overlap,
and the contamination is much lower because it scales as the square of the
intrinsic alignment amplitude $\propto C_2^2$ instead of linearly as in
the case of the $GI$ cross-correlation.

\section{Methods to assess the contamination}
\label{sec:methods}

So far we have considered the contamination of the power spectra in cosmic
shear surveys due to intrinsic alignments, and shown that the interference
between gravitational and intrinsic shears can dominate the contamination
and reach levels that are important for high precision studies of weak
lensing. In this section we discuss two methods to estimate the amplitude
of the effect and separate it from weak lensing effect.  The first method
is based on density-shear correlations, while the second method is purely
geometrical and uses only scaling of the cross-correlation signal with
redshift.

\subsection{Density-shear correlations}

We are assuming there is a correlation between 
the intrinsic alignment and gravitational potential 
field that gives rise to shear: the latter can be reconstructed  
from the density field. On large scales we can assume galaxy density 
field is linearly proportional to the density field up to a constant, 
the so called bias parameter. In fact, there is no need to do the 
reconstruction itself: at the two-point function level the full 
information on this correlation is encoded in the density-shear 
correlation.
We can thus constrain the intrinsic alignment models by comparing 
them to the observed density-shear correlation from galaxy surveys.  

To date these measurements have been done on small scales $<1h^{-1}$~Mpc
for the purpose of understanding contamination of galaxy-galaxy lensing.  
The quantity measured in these studies is $\Delta\gamma(r)$, defined as
the mean tangential shear of a ``satellite'' galaxy a transverse distance
$r$ from the ``primary.'' This is equal to
\begin{equation}
\Delta\gamma(r) = \frac{ \int \langle 
[1+\delta_g(0)][1+\delta_g(x)]\gamma_+^I(x)\rangle dx_\parallel }
  { \int \langle [1+\delta_g(0)]\delta_g(x)\rangle dx_\parallel },
\label{eq:a5}
\end{equation}
cf. Eq.~(A5) of Ref.~\cite{2004astro.ph..3255H}.  
The idea behind this equation is that we compute the galaxy-shear 
correlations for all close angular pairs on the sky and then use the 
corresponding galaxy-galaxy correlation to estimate what fraction of these
pairs is physically close also in three-dimensional space.  The signal is 
then boosted by dividing by this fraction. 

\begin{figure}
\includegraphics[angle=-90,width=3in]{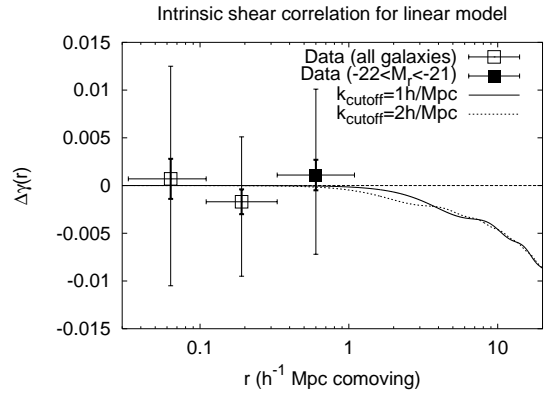}
\caption{\label{fig:dgr}The intrinsic shear statistic $\Delta\gamma(r)$ of
Eq.~(\ref{eq:a5}).  The SDSS data points are shown, with the horizontal
error bars indicating the range of radii used, the thick vertical error 
bars indicating $1\sigma$ statistical errors, and the thin vertical error 
bars representing 99.9\% confidence limits including systematics
(principally shear calibration and removal of lensing signal)
\cite{2004astro.ph..3255H}.}
\end{figure}

Noting that $(1+\delta_g)\gamma^I_+=\tilde\gamma^I_+$, and taking the 
Fourier transform in the transverse dimensions, we get
\begin{equation}
\Delta\gamma(r) = \frac{ \int P_{\delta_g,\tilde\gamma^I}(k) J_2(kr) k\, dk }
  { \int P_{\delta_g}(k) J_0(kr) k\, dk }
= \frac{ \int P_{\delta,\tilde\gamma^I}(k) J_2(kr) k\, dk }
  { b_g\int P_{\delta}(k) J_0(kr) k\, dk }.
\label{eq:dgr}
\end{equation}
In Fig.~\ref{fig:dgr} we have plotted $\Delta\gamma(r)$ as measured in the
Sloan Digital Sky Survey (SDSS) \cite{2004astro.ph..3255H};  the sample
was dominated by objects at $z\approx 0.1$ and so we have displayed the
model prediction at that redshift.  (The data points were presented in
Ref.~\cite{2004astro.ph..3255H} in physical separation units; we have
converted them to comoving separation at $z=0.1$.)

We see from Fig.~\ref{fig:dgr} that the model predicts a constant
value of density-shear correlations on large scales, as expected in a
linear bias model. On scales smaller than the initial size of a
galactic halo one does not expect much correlation.  The transition
scale is rather uncertain and depends on the somewhat arbitrary value
of the cutoff we used in the calculation. A typical value should be
given by the scale length within which the enclosed mass is of the
order of a typical galactic halo mass, of order $10^{12}M_{\odot}$.  
This gives $k_{\rm cutoff} \sim (1$--$2)h$/Mpc, which are the two
values used in Fig.~\ref{fig:dgr}. We see that the current constraints
are inconclusive because they do not extend to sufficiently large
scales.  The results from this figure also suggest that galaxy-shear
correlations are a powerful probe of identifying this contamination
and that there is plenty of statistical power in the SDSS data to
estimate the contamination by extending the analysis of
Ref.~\cite{2004astro.ph..3255H} to larger transverse separations.  

\subsection{Geometric projection}

The analysis of intrinsic alignment contamination should not be based
solely on low-redshift information: the local universe may not accurately
reflect the intrinsic alignments of the weak lensing source galaxies since
(for example) mergers could cause the intrinsic alignment signal to evolve
with redshift.  Consistency of shear cross-spectra obtained from different
types of galaxies in the same range of redshifts would argue against
significant gravitational-intrinsic correlations, but if an inconsistency
is observed that is not attributable to observational systematics, then
some method of separating pure lensing ($GG$) from gravitational-intrinsic
($GI$) signals will be needed.

One possibility is to take advantage of the different redshift dependence
in the $GG$ and $GI$ signals when applied to tomography between slices at
redshifts $z_\alpha<z_\beta$ (the redshift slices should be separated far
enough to eliminate overlaps in their distribution and hence $II$
contaminaiton).  In the idealized case where the photo-$z$s for the nearby
redshift bin $z_\alpha$ have negligible uncertainty, the $GI$ signal at a
fixed $\ell$ and $z_\alpha$ rises with increasing $z_\beta$ as
\begin{equation}
C_\ell^{EE,GI}(z_\alpha,\beta)\propto \int 
\frac{D_A(z_\alpha,z)}{D_A(0,z)}[1+\zeta(z)] \left({dn\over 
dz}\right)_\beta dz,
\label{eq:tom}
\end{equation}
where $D_A$ is the comoving angular diameter distance separating two
redshifts and $\zeta(z)$ is the shear calibration error for galaxies at
redshift $z$ (which ideally would be made negligible).  The $GG$ (pure
lensing) signal also rises with increasing $z_\beta$, but more slowly
because $z_\alpha$ in Eq.~(\ref{eq:tom}) is replaced by the redshift of
the contributing structures, which must be $<z_\alpha$.  In principle, one
could project out signals in tomography that have the redshift dependence
of Eq.~(\ref{eq:tom}).

If the effective ratio of angular diameter distances $g_{\alpha\beta}$ is
measured by cosmography -- i.e. measurement of the relative shear signal
at different $z_\beta$ behind a sample of lenses at $z_\alpha$ -- then
provided that the same source sample at $z_\beta$ is used for tomography
as for cosmography, the right-hand-side of Eq.~(\ref{eq:tom}) is obtained
directly \cite{2003PhRvL..91n1302J, 2004ApJ...600...17B} up to an
irrelevant multiplicative factor.  This geometrical method has the
practical advantage of automatically accounting for any shear calibration
errors, even if they are redshift-dependent, and eliminating the need for
external information about the angular diameter distance function $D_A$.  
Since the density-intrinsic shear correlation may evolve with redshift,
the de-projection of signals proportional to Eq.~(\ref{eq:tom}) will have
to be done separately for each redshift slice $z_\alpha$.  The method is
more accurate if the slices are narrow so that the effective $z_\alpha$
for the intrinsic alignment does not differ significantly from the
effective $z_\alpha$ for cosmography, thus precise photo-$z$s would be
extremely valuable.

The geometric method is model-independent in the
sense that it works for arbitrary $k$ and $z$-dependence of the
density-intrinsic shear correlation, but of course some cosmological
information is lost. An alternative approach would be to marginalize over
a parameterization of $P_{\delta,\tilde\gamma^I}(k,z)$ (perhaps forced to
match on to the $P_{\delta,\tilde\gamma^I}(k)$ measured in the local
universe at $z\approx 0$); this may retain more cosmological information
if $P_{\delta,\tilde\gamma^I}(k,z)$ is e.g. forced to be a smooth function
of $z$, but of course it sacrifices the model independence of the
geometric technique.

\section{Discussion}
\label{sec:discussion}

In this paper, we have considered the contamination of the power spectrum
in cosmic shear surveys due to intrinsic alignments, and shown that the
gravitational and intrinsic shears need not be independent.  For some
models, the contamination is dominated by the gravitational-intrinsic
cross-power rather than thet power spectrum of the intrinsic alignments
themselves.  One implication of this is that the $B$-mode signal,
frequently used as a systematics test, may be a misleading indicator of
intrinsic alignment contamination, since it is sensitive only to the
intrinsic-intrinsic autopower $C_\ell^{BB,II}$ but not to the
gravitational-intrinsic cross-power $C_\ell^{GI}$, which can only be
nonvanishing for the $E$ modes. Fig.~\ref{fig:broad}(c) provides an
extreme example: here the change in $C_\ell^{EE}$ due to intrinsic
alignments exceeds the $B$-mode power spectrum $C_\ell^{BB}$ by $>2$
orders of magnitude.  The opposite extreme is the quadratic alignment
model in which there is no gravitational-intrinsic correlation and the $E$
and $B$ mode contributions from intrinsic alignments are similar.

A realistic model of intrinsic alignments will likely fall somewhere in
between these two extremes.  The linear alignment model
(Eq.~\ref{eq:model1}) obviously predicts maximal correlation between the
density and intrinsic shear fields on large scales in the sense that the
correlation coefficient $\rho_{\delta,\tilde\gamma^I}(k)  =
P_{\delta,\tilde\gamma^I}(k)/\sqrt{P_{\delta}(k)P_{\tilde\gamma^I}(k)}$ 
has absolute value $\approx 1$.  Any higher-order corrections to
Eq.~(\ref{eq:model1}) will therefore reduce the correlation coefficient.  
On the other hand, the simplest form of the quadratic alignment model
(Eq.~\ref{eq:model2}) is probably too optimistic, even though the model is
well-motivated by tidal torque theory.  We found
$P_{\delta,\tilde\gamma^I}(k)=0$ for this model not because of any
fundamental symmetry principle, but rather because the assumed Gaussianity
of the density field causes third-order statistics in $\delta$ to vanish.  
In the real universe we can expect tidally torqued galaxies to have
$P_{\delta,\tilde\gamma^I}(k)\neq 0$ because of non-Gaussianity from
nonlinear evolution \cite{2002astro.ph..5512H}, and because of third- and
higher-order corrections to Eq.~(\ref{eq:model2}) that we have neglected.

While we have presented these results for the power spectrum, the
non-Gaussianity of the density field, and hence of cosmic shear, has
motivated studies of other statistics, most notably the bispectrum
\cite{2001ApJ...548....7C, 2001MNRAS.322..918V, 2002MNRAS.330..365H,
2003ApJ...598..818Z, 2004ApJ...607...40H, 2004MNRAS.348..897T}.  The
bispectrum can in principle pick up three types of intrinsic alignment
contamination: the intrinsic shear bispectrum ($III$), and the
gravitational-gravitation-intrinsic ($GGI$) and
gravitational-intrinsic-intrinsic ($GII$) cross-bispectra.  If tomography
is used, the measured bispectrum components are
$B_\labc^{EEE}(\alpha\beta\gamma)$, where $\alpha$, $\beta$, and $\gamma$
are indices indicating the source bins centered at redshifts $z_\alpha\le
z_\beta\le z_\gamma$.  Assuming that these bins do not overlap
significantly in redshift, the $III$ contaminant is only nonvanishing for
the triplets where all three source galaxies lie in the same redshift
slice $z_\alpha=z_\beta=z_\gamma$; this is only a small fraction of the
triplets and hence little statistical power is lost by rejecting them
\cite{2004ApJ...601L...1T}.  $GGI$ is produced by a
density-density-intrinsic shear bispectrum in the nearest of the three
slices, and it can in principle contaminate any tomographic shear
bispectrum in which $z_\alpha<z_\beta\le z_\gamma$.  $GII$ is produced by
the density-intrinsic-intrinsic bispectrum, again in the nearest of the
three slices, but this time in order to have the intrinsic shear be
correlated and to produce a lensing effect on the more distant slice,
$GII$ satisfies a ``selection rule'' $z_\alpha=z_\beta<z_\gamma$.  Thus
every triplet of source screens is potentially contaminated by exactly one
of $GGI$, $GII$, or $III$.  Note also that the quadratic alignment model
(Sec.~\ref{ss:m2}), which at lowest order predicted no $GI$ contamination
to the power spectrum, does predict a $GGI$ contamination
$B_{\delta,\delta,\tilde\gamma^I}(k_1,k_2,k_3)\neq 0$ because the 
intrinsic shear $\gamma^I$ is a quadratic function of $\delta$.  We defer 
a calculation of the magnitude of this effect to a future paper.

Our analysis clearly motivates observations to constrain the
density-intrinsic shear correlation.  Just as the intrinsic shear
autopower can be constrained by low-redshift measurements where the
lensing signal is negligible \cite{2000ApJ...543L.107P,
2002MNRAS.333..501B}, so it should be possible to use these same
low-redshift measurements to measure $P_{\delta,\tilde\gamma^I}(k)$
provided that a tracer of the density field is available.  The number
density of galaxies is one possibility; recent observations have probed
the galaxy-intrinsic shear correlation on $0.02$--$1 h^{-1}$~Mpc scales
\cite{2002AJ....124..733B, 2004astro.ph..3255H}, and it would be useful to
extend these studies to larger scales where linear biasing is valid.  (At
least one density-shear correlation measurement is available on large
scales \cite{2001ApJ...555..106L}.  These authors were interested
primarily in using intrinsic alignments to reconstruct the density field;
they therefore did not use lensing shear estimators, and they measured
three-dimensional separations in redshift space, so their results are
difficult to interpret in the present context.)  For such observations, it
would also be useful to distinguish types of galaxies (e.g. early vs.
late); if one type of galaxy turns out to have significantly greater
$P_{\delta,\tilde\gamma^I}(k)$, the shear power spectra could be performed
principally on the other types so as to reduce contamination.

A second approach to eliminate the contamination is to project out the
information that correlates with the known redshift scaling of the signal
in the cross-correlations between narrow redshift bins, assuming these are
available from the photo-$z$ methods.  This method is geometric and
achieves perfect decontamination under these idealized assumptions, but
projection does destroy some information in the data. It also puts severe
requirements on the accuracy of photometric redshifts, which may require
many passbands to achieve this \cite{2004astro.ph..3384B}.

In summary, we have shown that cosmic shear surveys will need to consider
the possibility of intrinsic alignment contamination from its interference
with weak lensing induced shear.  The magnitude of the effect is extremely
model-dependent and can range from essentially zero (simplest quadratic
alignment models) to severe (simplest linear alignment models).  In the
latter case it can exceed intrinsic correlations by an order of magnitude
and cannot be identified using $B$-mode power spectrum, nor can it be
eliminated using cross-correlations between different redshift slices.  
This result is disappointing and implies that this effect could be very
damaging for efforts to use weak lensing as a high precision test of
cosmology with future surveys such as Pan-STARRS, LSST or SNAP.  However,
the situation is far from hopeless: here we have considered several
possible ways to constrain and/or suppress the contamination.  The next
step observationally is a density-shear correlation analysis with the
shear computed using a lensing estimator in a wide-angle, shallow survey
such as the Sloan Digital Sky Survey \cite{2000AJ....120.1579Y} to
constrain $P_{\delta,\tilde\gamma^I}(k,z\ll 1)$; the next step
theoretically is a thorough analysis of the various methods described
above for suppressing the gravitational-intrinsic contamination, in
particular to understand how much degradation on the cosmological
parameters occurs and whether they increase sensitivity to any of the
other potential systematics such as photo-$z$ errors, shear calibration
errors, and spurious power.

\acknowledgments

C.H. acknowledges the support of the NASA Graduate Student Researchers
Program (GSRP).  U.S. is supported by Packard Foundation, NASA NAG5-11489
and NSF CAREER-0132953.

We thank R. Bean, S. Bridle, B. Joachimi, D. Kirk, and I. Laszlo for bringing an error in the original version of this article to our attention.

\end{document}